\documentclass[hidelinks]{gistt}

\usepackage{amsmath,amssymb,amsfonts}
\usepackage{algorithmic}
\usepackage{graphicx}
\usepackage{textcomp}
\usepackage{xcolor}
\usepackage{enumitem}

\bibliography{references.bib}

\title{Industry 4.0 Connectors - \\ A Performance Experiment with Modbus/TCP}
\author{Christian Nikolajew\\
Universität Hildesheim\\
Institut für Informatik\\
nikolajew@uni-hildesheim.de\\
31141 Hildesheim, Germany
\and
Holger Eichelberger\\
Universität Hildesheim\\
Institut für Informatik\\
eichelberger@sse.uni-hildesheim.de\\
31141 Hildesheim, Germany}

\begin{document}
\pagestyle{empty}

\newcommand{\etal}{~et~al.}
\newcommand{\IIPEcosphere}[0]{IIP-Ecosphere}
\newcommand{\OPCUA}[0]{OPC UA}
\newcommand{\oktoflow}[0]{oktoflow}
\newcommand{\idtaSpecs}[0]{IDTA-specs}
\newcommand{\idtaSpec}[0]{IDTA-spec}
\newcommand{\semanticId}[0]{semanticId}
\newcommand{\semanticIds}[0]{semanticIds}
\newcommand{\idShort}[0]{idShort}
\newcommand{\idShorts}[0]{idShorts}
\newcommand{\ModbusTcp}[0]{Modbus/TCP}

\maketitle

\begin{abstract}
For Industry 4.0 applications, communication protocols and data formats even for legacy devices are fundamental. In this paper, we focus on the \ModbusTcp{} protocol, which is, e.g., used in energy metering. Allowing Industry 4.0 applications to include data from such protocols without need for programming would increase flexibility and, in turn, improve development efficiency. As one particular approach, we discuss the automated generation of \ModbusTcp{} connectors for our Open Source \oktoflow{} platform and compare the performance of handcrafted as well as generated connectors in different settings, including industrial energy metering devices.

\end{abstract}

\section{Introduction}


Industry 4.0 applications integrate various devices, e.g., production machines, Programmable Logic Controller (PLC) or sensors. As the devices are often used for many years, shopfloors are usually heterogeneous and consist of devices of different age and version connected via by various protocols, e.g., OPC UA, MQTT, ADS~\cite{WeberEichelberger23} or, recently, Asset Administration Shell~\cite{SauerEichelberger23}. Still many PLCs, analog/digital input/output devices or energy meters are based on the \ModbusTcp{} protocol. For connecting devices by Industry 4.0 platforms, such as our Open Source \oktoflow{}\footnote{\url{https://oktoflow.de}} (former \IIPEcosphere{}) platform~\cite{EichelbergerNiederee23}, it is important to flexibly support a range of protocols. \oktoflow{} relies on a uniform connector layer which treats individual protocols as special cases. To reduce technical efforts, \oktoflow{} integrates protocols in a model-driven manner into applications turning generic connectors into application-specific connector instances.

In this paper, we ask the questions whether \ModbusTcp{} can also be addressed through our unifying connector layer, whether there are performance differences and whether a software server may realistically emulate a real device, e.g., for a digital twin. As contribution, we discuss our connector approach as well as the results of a performance experiment involving devices from Siemens and Phoenix Contact.

Some scientific work is available on the \ModbusTcp{} protocol. Recent work targets among other topics the security modeling and analysis of the protocol (e.g.~\cite{GOLDENBERG201363}), the performance analysis of the \ModbusTcp{} stack in~\cite{Bai_2018}, 
the scalability of the protocol depending on bit rate errors~\cite{9960319} as well as (hardware) integrations up to WiFi connections~\cite{9325535}. However, we are not aware of works on  an uniform model-driven integration and its performance effects.

\textit{Structure:} We touch in Section~\ref{sect-background} briefly the background of \ModbusTcp{}. In Section~\ref{sect-approach}, we introduce our generative approach, in Section~\ref{sect-eval} we discuss our performance experiment and in Section~\ref{sect-conclusion} we conclude.

\section{Background}\label{sect-background}

\ModbusTcp{} was invented in 1979 as open protocol for PLC communication and standardized in 2007 as part of IEC 61158. The protocol can be applied to various transmission media, e.g., as Modbus/RTU for serial communication via RS232, RS422 or RS485 via Ethernet (often 10/100 MBit/s) in terms of Modbus/TCP. We focus in this paper on Modbus/TCP.

A \ModbusTcp{} device offers digital data as individual bits of the \textit{discrete input} (read-only) and the read/write \textit{coils} space. For analogous data, 16 bit read-only \textit{input registers} and read-write \textit{holding registers} can be used. More complex types, e.g., floats or strings are stored in multiple 16 bit registers. In total, \ModbusTcp{} allows for $2^{16}$ data registers. However, the meaning as well as the byte order in the 16 bit registers are vendor dependent.

For example the EEM-MA370 energy meter by Phoenix Contact provides up to 57724 registers\footnote{\label{foot_eem}\url{https://www.phoenixcontact.com/de-de/produkte/energiemessgeraet-eem-ma370-2907983}}, including instantaneous values and statistical summaries of voltage (quality) or currency for three electrical phases. Based on practical demonstrator experience~\cite{EichelbergerNiederee23} and discussions with industrial colleagues, often an application obtains only some values, e.g., the effective power on individual phases.


\section{Integration Approach}\label{sect-approach}


\begin{figure}[tbp] 
\centerline{\includegraphics[page=1,trim={1.0cm 9.5cm 16.2cm 2.1cm},clip, scale=0.52]{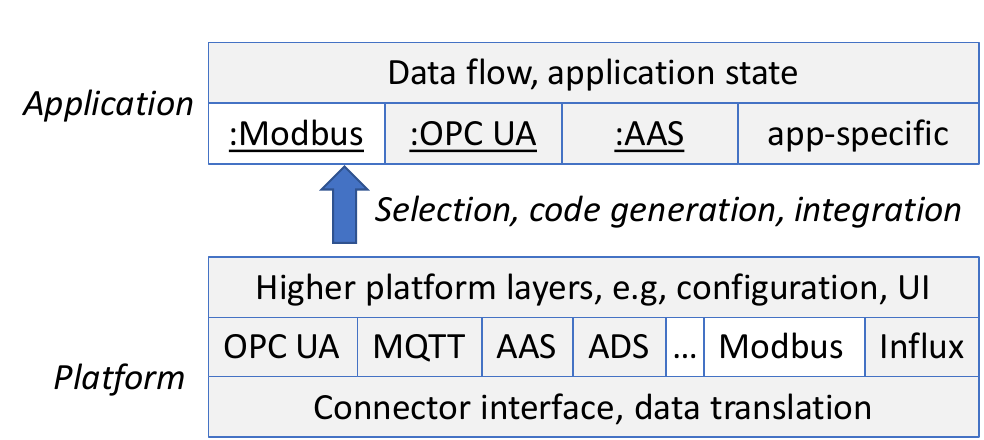}}
\caption{Connector types vs. instances in oktoflow.}
\label{fig_approach}
\end{figure}

At a glance, a \ModbusTcp{} connector consists of a protocol library and application-specific code to access individual values and to convert them taking vendor-defined types such as \texttt{uint16} or \texttt{float32} (akin to \cite{WeberEichelberger23}) and byte-order into account, etc. As Industry 4.0 applications typically involve various protocols, a developer would have to repeat the implementation exercise with different libraries and different data formats.

In the Open Source Industry 4.0 platform \oktoflow{}, we automate such development activities in model-driven manner. We distinguish between a generic protocol implementation realizing the \textbf{connector type} and the actual customized use in an application, the \textbf{connector instance} as illustrated in Figure~\ref{fig_approach}. The connector type must comply with platform connector interfaces encapsulating the access to individual fields, values or registers through (qualified) "names". The \oktoflow{} metamodel defines a connector as a "service", which translates external data into application data and back. External and internal data is specified in terms of (nested) typed fields. On model level, the \ModbusTcp{} connector specifies the byte order, prohibits nested data types and non-applicable field types through constraints and requires for each (external) data field a \ModbusTcp{} register offset. The application code generation reuses the generic connector and complements it with application-specific data converters, which ultimately access the data fields through the generic connector type.

As the core of \oktoflow{} is written in Java, we focus on a Java-based connector type. We initially examined 5 Java \ModbusTcp{} libraries for their last release, their documentation, their availability on Maven central and their expected performance. For the performance, we ran a test program that reads 400 values and writes 200 values from a \ModbusTcp{} software server and records the average results for 3 runs. In more details, the (total) access results were for Jamod $\approx{}$53ms, j2mod $\approx{}$57ms, JLibModbus $\approx{}$86ms, EasyModbus $\approx{}$661ms and Modbus4j $\approx{}$8541ms. We selected Jamod\footnote{\url{https://jamod.sourceforge.net/}} for the implementation of our polling-based connector type and validated the realization with the Jamod software server.

\section{Experiments}\label{sect-eval}


To answer our questions, we performed experiments to determine selected runtime properties.

\textbf{Environment:} Software-based measurements executed locally on a laptop, in a local network with 100 MBit/s running connector and server on two computers as well as measurements with real devices, a Siemens Sentron PAC4200\footnote{\url{https://cache.industry.siemens.com/dl/files/595/34261595/att_951629/v1/manual_pac4200_de-DE_de-DE.pdf}} in a 100 MBit/s network and a Phoenix Contact EEM-MA370$^{\ref{foot_eem}}$ with 100 MBit/s interface in a 1000 MBit/s network, whereby the devices are part of setups that shall not be modified. As laptop, we used a Lenovo ideapad 330 with i5-8250U@1.6GHz, 8GB RAM and JDK 15.0.2.

\textbf{Subjects:} Handcrafted (EEM) and generated (EEM, Sentron) connectors based on the Jamod library, set up for 10 (mostly \texttt{float32}) values. For further insights, emulating the approach of our colleagues from mechanical engineering, we used a handcrafted Python script on Python 3.8.10 reading the same values through pyModbusTCP 0.2.1.

\textbf{Procedure:} Due to the construction of our connectors, we measure and log the time required to read the 10 values as one batch. In an experiment step, we read 5000 batches and archive the log. For each  software-based connector measurement, we collected 18k data points, for a device connector run 12k and for an indicative Python script execution on a device 6k data points. We perform a pre-experiment to identify issues or when Java JIT settles. Ultimately, we repeat the experiment step three times to handle deviations or outliers.

\textbf{Results\footnote{\url{https://doi.org/10.5281/zenodo.13902632}}:} The pre-experiments indicated that Java stabilizes around 1500 batches, while there is no settling time for Python. However, we uniformly remove the first 1500 batches from all final measurements. When we ran the Java connectors the first time on the real devices, we surprisingly received no data. As a root cause, we identified the jamod library, which we could easily replace by its fork j2mod\footnote{\url{https://github.com/steveohara/j2mod}}. The received data requires individual byte order configurations as the Sentron works with big endian while the EEM with small endian.

\begin{table*}[ht]
\begin{center}
\begin{tabular}{c|c|c|c|c|c|c|c|c|c|c}
     \hline
     [$\mu{}s$]        & \multicolumn{3}{c|}{local, no network} & \multicolumn{3}{c|}{local, network}& \multicolumn{4}{c}{devices, network}\\
             & manual & EEM & Sentron     & manual & EEM & Sentron            & EEM   & EEM-py & Sentron & Sentron-py\\
     \hline
     Min     & 632  & 621  & 581          & 1858  & 1899  & 1928              & 6023  & 2724   & 2653    & 106\\
     Avg     & 1290 & 1295 & 1303         & 4802  & 4349  & 4709              & 13233 & 7907   & 6760    & 1277\\
     Median  & 1275 & 1280 & 1287         & 4924  & 4368  & 4672              & 12453 & 5961   & 4336    & 480\\
     Max     & 5551 & 5491 & 4837         & 10860 & 14280 & 9182              & 66610 & 49822  & 30219   & 22106\\
     Stddev  & 139  & 137  & 138          & 1510  & 1363  & 1559              & 6586  & 5420   & 5842    & 2977\\
     \hline
\end{tabular}
\label{tab_results}
\caption{Average results for software-based and device-based measurements (settling time removed).}
\end{center}
\end{table*}

Table~\ref{tab_results} summarizes the average results for all measurements, i.e., the manual EEM connector and the generated connectors in all three environments. In the local environment (no network), the measurements do not differ significantly. Linking connector and software server through network on two computers, the average response time increases by around factor 3.5 and the standard deviation by factor 10.5. The generated EEM connector appears to be slightly faster than the Sentron or the manual connector. 

For the two employed devices, the measurements differ significantly. The results for the EEM are about factor 2 higher than the results for the Sentron, while the standard deviation does not differ much. These differences may be due to the hardware of the devices (a Sentron is factor 6 more expensive than an EEM). As the request cycle for a single value takes roughly 0.7 ms on the Sentron and 1,3 ms on the EEM, we can obtain about 770-1500 measurements per second. This speed might not even be needed as the devices provide statistical summaries of their measurements, i.e., a lower polling frequency for already aggregated values may be feasible in an application.

For the Python scripts we found that the average response times for the EEM are about factor 2 lower than for the Java connector while the standard deviation does not differ much. For the Sentron, the Python script is about factor 5 faster and the standard deviation drops by factor 2. As the environment and the devices did not change, the difference is probably due to a more efficient library implementation.

\textbf{Limitations:} The results presented in this paper stem from a systematic experimentation with a small set of devices in a specific software environment (oktoflow), i.e., with limited generalizability. Still, our results give an impression about the expected performance and even indicate surprising differences, e.g., among the utilized devices. 

\section{Conclusion}\label{sect-conclusion}


In this paper, we presented a flexible, low-effort integration of the \ModbusTcp{} protocol into a software platform based on uniform interfaces and model-driven generation. We analyzed handcrafted and generated connectors in a performance experiment involving two industry-grade energy measurement devices. While the unified approach is feasible, the performance of the devices and of the analyzed software libraries differs, e.g., Java vs. Python. Based on our observations, we do not recommend using a software-based \ModbusTcp{} server as an easy-to-realize digital twin for a real device without extensive experiments, as either protocol functionality or performance may differ significantly and additional components could be needed to simulate the realistic behavior. 

In future work, we plan to modify our code generation so that primitive data types can be passed without additional type casts, which might improve the performance and evaluate the generated connectors in real-world application scenarios.

\section*{Acknowledgement}
We would like to thank Dr.~Thomas Lepper and Aleks Arzer (Institut für Fertigungstechnik und Werkzeugmaschinen, Garbsen) for supporting our experiments on their premises as well as Thomas Hildebrandt (Phoenix Contact) for background information.

\printbibliography

\end{document}